\documentclass[12pt]{iopart}
\usepackage{iopams}

\usepackage{amssymb}
\usepackage{graphicx,psfrag,color}
\usepackage{dcolumn}
\usepackage{bm}
\usepackage{mathbbol}

\begin{document}


\title{A Simple Derivation of Schr\"odinger Uncertainty Relation}

\author{Gustavo Rigolin}
\address{Departamento de F\'isica, Universidade Federal de
S\~ao Carlos, 13565-905, S\~ao Carlos, SP, Brazil}
\ead{rigolin@ufscar.br}

\date{\today}

\begin{abstract}
We show how the Schr\"odinger Uncertainty Relation for a pair of
observables can be deduced using the Cauchy-Schwarz inequality
plus successive applications of the commutation relation involving
the two observables. Our derivation differs from the original one
in the sense that we do not need the expansion of the product of
these two observables in a sum of symmetrical and anti-symmetrical
operators.
\end{abstract}


\vspace{2pc}


\section{Introduction}

In 1930 Erwin Schr\"odinger presented \cite{sch30,bat99} a lower
bound for the product of the dispersion of two non-commuting
observables. This lower bound, from now on called Schr\"odinger
Uncertainty Relation (SUR), is more general than the usual
Heisenberg Uncertainty Relation (HUR) taught in all quantum
mechanics courses. In fact, HUR can be
derived from SUR while SUR does not follow from HUR.

In this article we present an alternative derivation of
Schr\"odinger's relation. Different from the original one, we do
not make use of the expansion of the product of two observables in
a sum of symmetrical and anti-symmetrical operators.

\section{Schr\"odinger's derivation}
\label{sch}

Let $X$ and $P$ be our two non-commuting observables such that $[X,P]=XP-PX
\neq 0$. Now we define the following two states,
\begin{equation}
|\psi\rangle = X |\chi \rangle
\label{psi}
\end{equation}
and
\begin{equation}
|\phi\rangle = P |\chi \rangle. \label{phi}
\end{equation}
For the moment we assume $\langle X \rangle = \langle \chi | X
|\chi \rangle=0$ and $\langle P \rangle = \langle \chi | P |\chi
\rangle=0$. These quantities are, respectively, the mean values of
$X$ and $P$ for a system described by the normalized state
$|\chi\rangle$. Applying the Cauchy-Schwarz inequality for the
states (\ref{psi}) and (\ref{phi}),
\begin{equation}
\langle \psi | \psi \rangle \langle \phi | \phi \rangle \geq
|\langle \psi | \phi \rangle|^2 = \langle \psi | \phi \rangle
\langle \phi | \psi \rangle, \label{schwarz}
\end{equation}
we get
\begin{equation}
\left\langle X^2 \right\rangle \left\langle P^2 \right\rangle \geq
\langle X P \rangle\langle P X \rangle. \label{step1}
\end{equation}
Remembering that the dispersions of $X$ and $P$ are $\Delta
X=(\langle X^2 \rangle -\langle X \rangle^2)^{1/2}$ and $\Delta
P=(\langle P^2 \rangle -\langle P \rangle^2)^{1/2}$,
Eq.~(\ref{step1}) becomes
\begin{equation}
(\Delta X)^2 (\Delta P)^2 \geq \langle X P \rangle\langle P X
\rangle. \label{step2}
\end{equation}
Now we make use of Schr\"odinger's ingenuity and write $XP$ as a
sum of symmetrical (S) and anti-symmetrical (A) operators:
\begin{eqnarray}
XP &=& \frac{XP + PX}{2} + \frac{XP - PX}{2} = S + A, \label{s} \\
PX &=& \frac{XP + PX}{2} - \frac{XP - PX}{2} = S - A. \label{a}
\end{eqnarray}
Inserting Eqs.~(\ref{s}) and (\ref{a}) in Eq.~(\ref{step2}) we obtain
\begin{eqnarray}
(\Delta X)^2 (\Delta P)^2 &\geq & \langle S + A \rangle\langle S -
A \rangle \nonumber \\
& \geq & \langle S \rangle^2 - \langle A \rangle^2 = \frac{\langle \{ X, P\}\rangle^2}{4} - \frac{\langle [ X,
P]\rangle^2}{4}, \label{final1}
\end{eqnarray}
where $\{ X, P\}=XP + PX$ is the anti-commutator of $X$ and $P$.
Moreover, since $X$ and $P$ are Hermitian operators (observables)
we know that $[X,P]=\mathrm{i}C$, where $C$ is Hermitian and
$\mathrm{i}=\sqrt{-1}$. Therefore, Eq. (\ref{final1}) can be
written as,
\begin{equation}
(\Delta X)^2 (\Delta P)^2 \geq \frac{\langle \{ X,
P\}\rangle^2}{4} + \frac{\left|\langle [ X,
P]\rangle\right|^2}{4}. \label{final2}
\end{equation}
Finally, if we had $\langle X \rangle \neq 0$ and $\langle P
\rangle \neq 0$ the proof can be carried out by making the
following substitutions:
\begin{eqnarray}
X & \longrightarrow & X - \langle X \rangle, \label{sub1}\\
P & \longrightarrow & P - \langle P \rangle. \label{sub2}
\end{eqnarray}
Repeating the previous procedure we arrive at the general form of
SUR:
\begin{equation}
(\Delta X)^2 (\Delta P)^2 \geq \left( \frac{\langle \{ X,
P\}\rangle}{2} - \langle X \rangle \langle P \rangle \right)^2 +
\frac{\left|\langle [ X, P]\rangle\right|^2}{4}. \label{sur}
\end{equation}

Note that HUR, namely,  $(\Delta X)^2 (\Delta P)^2 \geq \left|\langle [ X, P]\rangle\right|^2/4$,
follows from Eq. (\ref{sur}) if we drop the first term at the right hand side.

\section{An alternative derivation}
\label{new}

We now present a new way of deriving Eqs.~(\ref{final2}) and
(\ref{sur}) without employing Eqs.~(\ref{s}) and (\ref{a}), while
maintaining the same simplicity of the previous derivation. The
key idea behind the following deduction lies on the convenient use
of the commutator of $X$ and $P$.

Again, we begin with the simplest situation, i.e., $\langle
 X \rangle = \langle P \rangle = 0$ and, as before, our starting
 point is Eq.~(\ref{step2}). Remembering the definition of the
 commutator of $X$ and $P$ we can write $XP$ and $PX$ as
 \begin{eqnarray}
 XP &=& PX + [X,P], \label{com1} \\
 PX &=& XP - [X,P]. \label{com2}
 \end{eqnarray}
Using Eq.~(\ref{com1}) we can write Eq.~(\ref{step2}) as
\begin{equation}
(\Delta X)^2 (\Delta P)^2 \geq \langle P X \rangle^2+ \langle
[X,P] \rangle \langle P X \rangle. \label{2of4}
\end{equation}
Now using Eq.~(\ref{com2}) in Eq.~(\ref{step2}) we get
\begin{equation}
(\Delta X)^2 (\Delta P)^2 \geq \langle X P \rangle^2- \langle
[X,P] \rangle \langle X P \rangle. \label{3of4}
\end{equation}
Finally, using simultaneously Eqs.~(\ref{com1}) and (\ref{com2})
in Eq.~(\ref{step2}) we obtain,
\begin{eqnarray}
(\Delta X)^2 (\Delta P)^2 &\geq & \langle X P \rangle \langle P X
\rangle + \langle [X,P] \rangle \langle X P \rangle \nonumber \\
&  & - \langle [X,P] \rangle \langle P X \rangle - \langle [X,P]
\rangle^2. \label{4of4}
\end{eqnarray}
To finish the proof we add Eqs.~(\ref{step2}), (\ref{2of4}),
(\ref{3of4}), and (\ref{4of4}), which gives
\begin{equation}
(\Delta X)^2 (\Delta P)^2 \geq  \frac{\left( \langle X P \rangle
+ \langle P X \rangle\right)^2}{4} - \frac{\langle [X,P]
\rangle^2}{4}. 
\end{equation}
This equation can be written as 
\begin{equation}
(\Delta X)^2 (\Delta P)^2 \geq\frac{\langle \{ X, P\}\rangle^2}{4} + \frac{\left|\langle
[ X, P]\rangle\right|^2}{4}
\end{equation}
if we remember that $X$ and $P$ are hermitian operators. 
Finally, if $\langle X \rangle \neq 0$ and $\langle P \rangle \neq 0$ the
previous proof also works and we get SUR given by Eq.~(\ref{sur})
if we repeat the previous procedure using Eqs.~(\ref{sub1}) and
(\ref{sub2}).


\ack
The author thanks Funda\c{c}\~ao de Amparo \`a Pesquisa do Estado
de S\~ao Paulo (FAPESP) for funding this research.


\section*{References}

\begin{thebibliography}{25}

\bibitem{sch30} Schr\"odinger E 1930 
Zum Heisenbergschen Unsch\"arfeprinzip
\textit{Proceedings of The Prussian
Academy of Sciences} \textbf{19} 296-303. 
This text is written in german.

\bibitem{bat99} Angelow A and Batoni M C 1999 
About Heisenberg uncertainty relation (by E.Schrodinger) 
\textit{Bulg. J. Phys.} \textbf{26} 193-203 (1999). Also available at e-print:
quant-ph/9903100. This article is an english translation of Ref.
\cite{sch30}.


\end{thebibliography}
\end{document}